\documentclass[aip,reprint,showpacs,twocolumn,superscriptaddress,floatfix]{revtex4-1}
\usepackage{epsfig}
\usepackage[T1]{fontenc}
\usepackage[utf8]{inputenc}
\usepackage{amsmath}
\usepackage{amssymb}
\usepackage{amsfonts}
\usepackage{mathptmx}
\usepackage{dcolumn}
\usepackage{eucal}
\usepackage{bm}
\usepackage{color}
\usepackage[colorlinks,linkcolor=blue,citecolor=blue]{hyperref}

\usepackage{epstopdf}
\begin{document}



\title{Electronic heat current rectification in hybrid superconducting devices}

\author{Antonio Fornieri}
\email{antonio.fornieri@sns.it}
\affiliation{NEST, Istituto Nanoscienze-CNR and Scuola Normale Superiore, I-56127 Pisa, Italy}

\author{María José Martínez-Pérez}
\affiliation{Physikalisches Institut - Experimentalphysik II Universit\"at T\"ubingen, D-72076 T\"ubingen, Germany}

\author{Francesco Giazotto}
\email{francesco.giazotto@sns.it}
\affiliation{NEST, Istituto Nanoscienze-CNR and Scuola Normale Superiore, I-56127 Pisa, Italy}




\begin{abstract}
In this work, we review and expand recent theoretical proposals for the realization of electronic thermal diodes based on tunnel-junctions of normal metal and superconducting thin films. Starting from the basic rectifying properties of a single hybrid tunnel junction, we will show how the rectification efficiency can be largely increased by combining multiple junctions in an asymmetric chain of tunnel-coupled islands. We propose three different designs, analyzing their performance and their potential advantages. Besides being relevant from a fundamental physics point of view, this kind of devices might find important technological application as fundamental building blocks in solid-state thermal nanocircuits and in general-purpose cryogenic electronic applications requiring energy management. 
\end{abstract}

\pacs{}

\maketitle


\section{Introduction}
A thermal rectifier \cite{Starr,LiPRL,RobertsRev} can be defined as a device that connects asymmetrically two thermal reservoirs: the heat current transmitted through the diode depends on the sign of the temperature bias imposed to the reservoirs. This non-linear device represents the thermal counterpart of the well-known electric diode, which helped the extraordinary evolution of modern electronics together with the transistor. The realization of efficient thermal rectifiers would represent a giant leap for the control of heat currents at the nanoscale, \cite{GiazottoRev,Dubi,LiRev} boosting emerging fields such as coherent caloritronics,\cite{GiazottoNature,MartinezNature,MartinezRev} nanophononics,\cite{LiRev} and thermal logic.\cite{LiRev} Furthermore, a great number of nanoscience fields, including solid state cooling,\cite{GiazottoRev} ultrasensitive cryogenic radiation detection\cite{GiazottoRev,GiazottoHeikkila} and quantum information,\cite{NielsenChuang,Spilla} might strongly benefit from the possibility of releasing the dissipated power to the thermal bath in an unidirectional way.  

A highly efficient thermal diode should provide differences of at least one order of magnitude between the heat current transmitted in the forward temperature-bias configuration, $J_{\rm fw}$, and that generated upon temperature bias reversal, $J_{\rm rev}$. This is equivalent to say that the rectification efficiency
\begin{equation}
\mathcal{R}=\frac{J_{\rm fw}}{J_{\rm rev}}
\end{equation} 
must be $\gg 1$ or $\ll 1$.

Since the first theoretical proposal,\cite{LiPRL} several groups have put a great effort into envisioning different designs for thermal rectifiers dealing with phonons,\cite{Terraneo,Li,Segal,Segal2} electrons\cite{SanchezPRB,Ren,Ruokola1,Kuo,Ruokola2,Chen} and photons.\cite{BenAbdallah} Alongside these theoretical works, promising experimental results were obtained in systems that exploited phononic\cite{Chang,Kobayashi,Tian} and electronic\cite{Scheibner} thermal currents. Nevertheless, a maximum $\mathcal{R}\sim 1.4$ has been reported\cite{Kobayashi} and more efficient rectification mechanisms are required to realize competitive thermal diodes.

In this article, we will focus on the rectification performance of devices based on tunnel-junctions between normal metal (N) and superconducting (S) thin films at low temperatures. First, we will review the intrinsic properties of single NIN, NIS and S$_1$IS$_2$ junctions\cite{MartinezAPL,GiazottoBergeret} (where I stands for an insulating barrier and S$_1$ and S$_2$ represent two different superconducting electrodes). Then, we will analyze possible improvements of the rectification efficiency when these structures are combined together forming an asymmetric chain of tunnel-coupled islands.\cite{FornieriAPL,MartinezArxiv} In particular, as we shall argue, an asymmetric coupling to the phonon bath can strongly enhance the efficiency of the device, providing outstanding values of $\mathcal{R}$ with realistic parameters.
Very recently, the latter mechanism has been experimentally demonstrated in a hybrid device exhibiting a maximum rectification efficiency of $\sim 140$.\cite{MartinezArxiv}

\section{NIN and NIS junctions}\label{NISsec}
\begin{figure*}[t]
\centering
\includegraphics[width=0.6\textwidth]{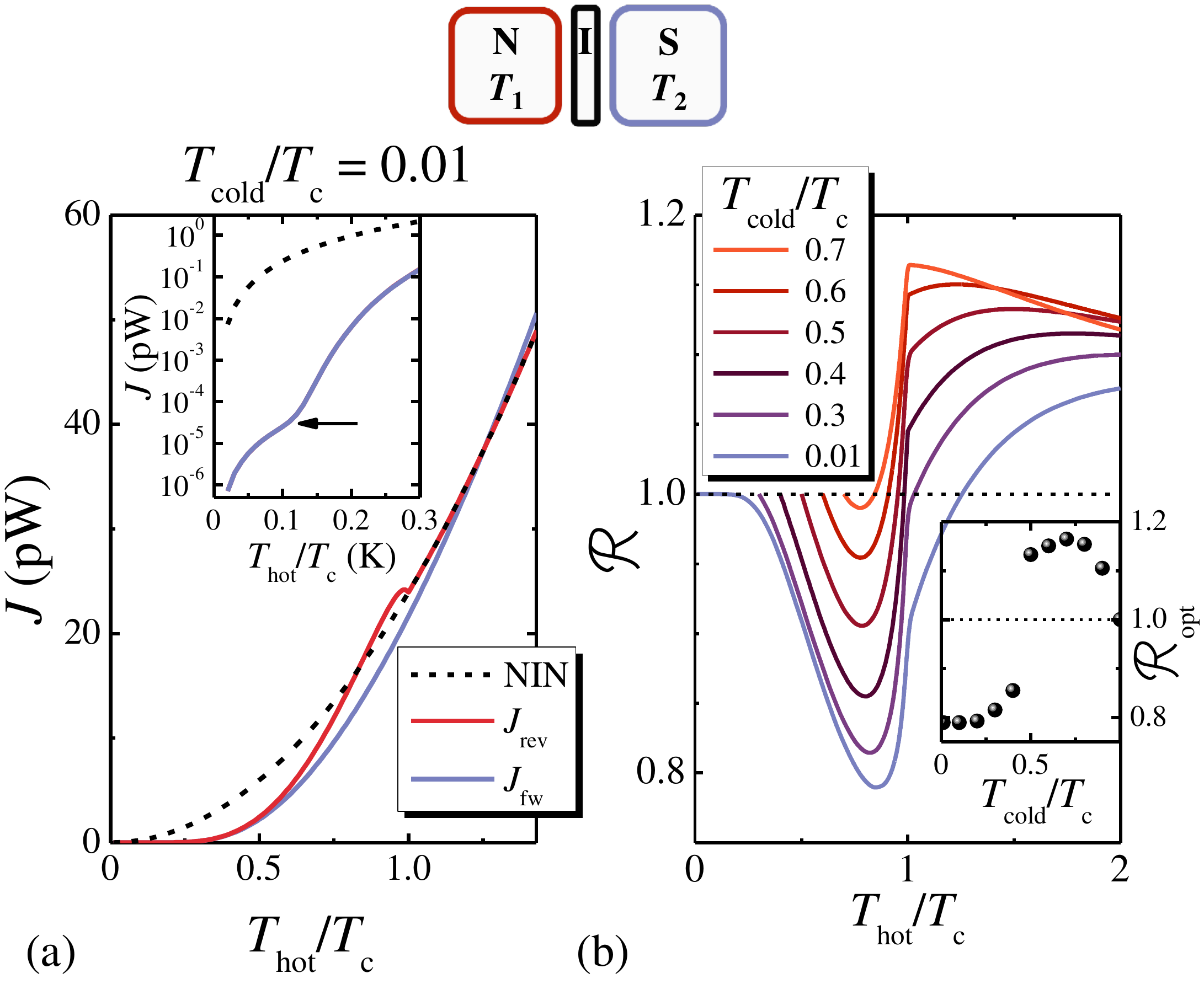}
\caption{(a) Thermal current $J$ vs. $T_{\rm hot}$ flowing through a NIN junction (dashed line) and through a NIS junction (solid lines). The curves are calculated for $R_{\rm T}=1$ k$\Omega$, $T_{\rm cold}=0.01 T_{\rm c}$ and $T_{\rm c}=1.4$ K. The inset highlights different trends of $J$ in the same cases as those displayed in the main panel for small values of $T_{\rm hot}$ on a logarithmic scale. (b) Rectification efficiency $\mathcal{R}$ vs. $T_{\rm hot}$ of a NIS junction for different values of $T_{\rm cold}$. Inset: optimal rectification efficiency $\mathcal{R}_{\rm opt}$ vs. $T_{\rm cold}$. Horizontal dashed lines indicate $\mathcal{R}=1$.}
\label{Fig1}
\end{figure*} 
We shall start, first of all, by describing the equations governing heat transport in simple NIN and NIS junctions. In these opening sections, we focus on the electronic heat currents only, and neglect possible contributions by lattice phonons. The latter will be discussed in the analysis of chain diodes. 

If we consider two N electrodes residing at electronic temperatures $T_{\rm hot}$ and $T_{\rm cold}$ (with $T_{\rm hot}\geq T_{\rm cold}$) coupled by means of a tunnel junction, the stationary electronic thermal current flowing through the junction can be written as:\cite{GiazottoBergeret} 
\begin{equation}
J_\mathrm{NIN}(T_{\rm hot},T_{\rm cold})=\frac{k_{\mathrm{B}}^2 \pi^2}{6e^2 R_\mathrm{T}}(T_{\rm hot}^2-T_{\rm cold}^2),\label{JNIN}
\end{equation}
where $R_\mathrm{T}$ is the contact resistance, $e$ is the electron charge and $k_\mathrm{B}$ is the Boltzmann's constant. Equation (\ref{JNIN}) clearly shows that no rectification is possible in a full normal-metal tunnel junction, since $\vert J_\mathrm{NIN}(T_{\rm hot},T_{\rm cold})\vert = \vert J_\mathrm{NIN}(T_{\rm cold},T_{\rm hot})\vert$.

However, if we substitute the second N electrode with a superconductor, the thermal current flowing through the NIS junction becomes:\cite{GiazottoRev,MakiGriffin}
\begin{eqnarray}
\begin{aligned}
J_\mathrm{NIS}(T_{\rm hot},T_{\rm cold})=\frac{2}{e^2 R_{\rm T}} \int_0^{\infty} & dE E \mathcal{N}(E,T_{\rm cold})\\
\times &[f(E,T_{\rm hot})-f(E,T_{\rm cold})].\label{JNIS}
\end{aligned}
\end{eqnarray}
Here, $ \mathcal{N}(E , T)=\left| \Re \left[  E+i \Gamma/ \sqrt{(E+i \Gamma)^2- \Delta^2(T)} \right] \right|$ is the smeared (by non-zero $\Gamma$) normalized Bardeen-Cooper-Schrieffer (BCS) density of states (DOS) in the superconductor,\cite{Dynes,PekolaPRL2004,PekolaPRL2010} $\Delta(T)$ is the temperature-dependent superconducting energy gap with a critical temperature $T_{\rm c}=\Delta(0)/(1.764 k_{\rm B})$ and $f(E,T)=[1+\textrm{exp}(\frac{E}{k_{\rm B}T})]^{-1}$ is the Fermi-Dirac distribution function. In the following, we will assume $\Gamma \sim 10^{-4}\Delta(0)$, which describes realistic NIS junctions.\cite{MartinezArxiv,PekolaPRL2004,PekolaPRL2010} We also define $J_{\rm fw}=\vert J_{\rm NIS}(T_{\rm hot},T_{\rm cold})\vert$ and $J_{\rm rev}=\vert J_{\rm NIS}(T_{\rm cold},T_{\rm hot})\vert$ for the forward and the reverse configuration, respectively. Figure \ref{Fig1}(a) shows the influence of the superconducting DOS on $J$: at a first glance, $J_{\rm fw}$ immediately appears different from $J_{\rm rev}$, thanks to the temperature dependence of $\Delta(T)$, which breaks the thermal symmetry of the tunnel junction. 
For $T_{\rm hot}>0.4T_{\rm c}$ the value of $\Delta(T)$ in the reverse configuration starts to decrease and $J_{\rm rev}$ gets significantly larger than $J_{\rm fw}$, leading to $\mathcal{R}<1$, as shown in Fig. \ref{Fig1}(b). On the contrary, if $T_{\rm hot}$ is raised above $T_{\rm c}$, heat flux from N to S becomes preferred, resulting in $\mathcal{R}>1$. As $T_{\rm cold}$ is increased, the reverse rectification regime gets gradually suppressed, while the forward regime becomes more efficient.   
We can highlight this behavior by introducing the optimal rectification efficiency $\mathcal{R}_{\rm opt}$, defined as that corresponding to the maximum value between $\mathcal{R}$ and $1/\mathcal{R}$ at a given $T_{\rm cold}$. This quantity provides the best working point of the diode, which can occur in the forward (if $\mathcal{R}_{\rm opt}>1$) or in the reverse (if $\mathcal{R}_{\rm opt}<1$) regime of rectification. As displayed in the inset of Fig. \ref{Fig1}(b), for $T_{\rm cold}<0.5 T_{\rm c}$ the reverse regime is the most efficient, while at higher temperatures the forward regime is favored. Finally, at $T_{\rm cold}=T_{\rm c}$ the system turns into a NIN junction and no rectification is possible.

It is also worth noting that at low values of $T_{\rm hot}$ and $T_{\rm cold}$, $J_{\rm NIS}$ is suppressed by a factor $\Gamma$ with respect to $J_{\rm NIN}$ and does not depend on the sign of the temperature bias [see inset of Fig. \ref{Fig1}(a)], indicating the effectiveness of $\Delta(T)$ as a thermal bottleneck. On the other hand, when $T_{\rm hot}$ and $T_{\rm cold}$ increase, $f(E,T_{\rm hot})-f(E,T_{\rm cold})$ becomes significantly different from zero at energies higher than the smeared $\Delta(T)$: this corresponds to a rapid increase of both $J_{\rm fw}$ and $J_{\rm rev}$, as indicated by the arrow in the inset. Thus, the thermal bottleneck effect generated by the superconducting energy gap decays quickly by raising temperatures. Although this change of behavior is not relevant for the rectifying properties of a single NIS junction, it will be important to understand the operation of a NINISIN chain. 
\begin{figure*}[t]
\centering
\includegraphics[width=0.6\textwidth]{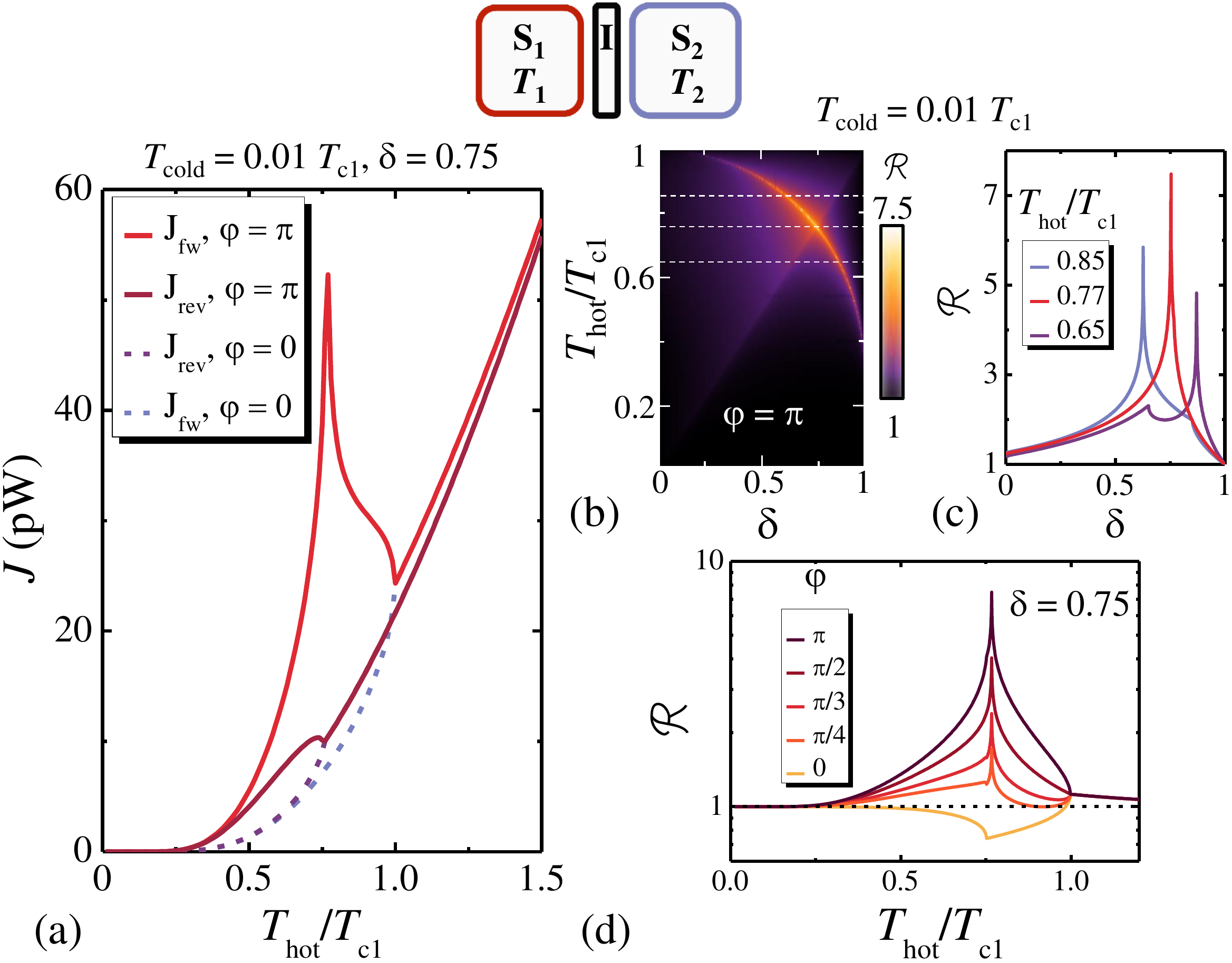}
\caption{(a) Heat currents $J_{\rm fw}$ and $J_{\rm rev}$ vs. $T_{\rm hot}$ flowing through a S$_1$IS$_2$ junction for $\varphi=0,\pi$. The curves are calculated for $\delta=0.75$, $R_{\rm T}=1$ k$\Omega$, $T_{\rm cold}=0.01 T_{\rm c1}$ and $T_{\rm c1}=1.4$ K. (b) Contour plot showing $\mathcal{R}$ as a function of $\delta$ and $T_{\rm hot}$ for $\varphi=\pi$ and $T_{\rm cold}=0.01 T_{\rm c1}$. (c) Three selected profiles of $\mathcal{R}$ vs. $\delta$ for the values of $T_{\rm hot}$ corresponding to the dashed lines in panel (b). (d) $\mathcal{R}$ vs. $T_{\rm hot}$ for $\delta=0.75$ and for different values of $\varphi$ at $T_{\rm cold}=0.01 T_{\rm c1}$. The horizontal dashed line indicates $\mathcal{R}=1$.}
\label{Fig2}
\end{figure*}
\section{S$_1$IS$_2$ junctions}
We now move to the case in which both the tunnel-coupled electrodes are superconductors, i.e., a S$_1$IS$_2$ junction.\cite{MartinezAPL} The latter can be thermally biased by setting the quasiparticle temperature of one electrode to $T_{\rm hot}$ and that of the other to $T_{\rm cold}$.  Moreover, we assume that S$_1$ and S$_2$ are characterized by different superconducting energy gaps $\Delta_1$ and $\Delta_2$ and by a phase difference $\varphi=\varphi_2-\varphi_1$. Then, the forward and reverse total heat currents flowing through the Josephson junction read:\cite{MakiGriffin,GiazottoAPL12}
\begin{eqnarray}
\begin{aligned}
 J_{\rm S_1IS_2,fw(rev)}=\;\; & J_{\rm qp}[ T_{\textrm{hot(cold)}}, T_{\textrm{cold(hot)}}] \\
 &-J_{\rm int}[T_{\textrm{hot(cold)}}, T_{\textrm{cold(hot)}}]\cos\varphi. \label{JSIS} 
\end{aligned}
\end{eqnarray}
Here, the term $J_{\rm qp}$ accounts for the energy carried by quasiparticles\cite{MakiGriffin,Golubev,Guttman,Guttman2,Zhao1,Zhao2} and is equivalent to Eq. \ref{JNIS} for the present system:
\begin{eqnarray}
\begin{aligned}
J_{\rm qp}(T_{\rm hot}, T_{\rm cold}) = \frac{ 2 }{e^2 R_{\textrm{T}} } \int_0 ^\infty dE E  \mathcal{N}_1(&E, T_{\rm hot}) \mathcal{N}_2  (E, T_{\rm cold})\\ \times &  [f(T_{\rm hot}) - f(T_{\rm cold}) ],
\label{Jquasiparticles} 
\end{aligned}
\end{eqnarray}
where $\mathcal{N}_1$ and $\mathcal{N}_2$ are the smeared normalized superconducting DOS of S$_1$ and S$_2$, respectively. On the other hand, $J_{\rm int}$ is the phase-dependent component of the heat current:\cite{MakiGriffin,Guttman,Guttman2,Zhao1,Zhao2}
\begin{eqnarray}
\begin{aligned}
J_{int}(T_{\rm hot}, T_{\rm cold}) = \frac{ 2}{e^2 R_{\textrm{T}} }    \int _0 ^\infty dE E \mathcal{M}_1(&E , T_{\rm hot}) \mathcal{M}_2   (E , T_{\rm cold}) \\ \times & [f(T_{\rm hot}) - f(T_{\rm cold}) ],
\label{int} 
\end{aligned}
\end{eqnarray}
where  $\mathcal{M}_{k}(E , T) = \left| \Im  \left[ -i\Delta_{k}(T)/\sqrt{ (E+i\Gamma_{k})^2 -\Delta_{k}^2(T)}  \right] \right|$ is the Cooper pair BCS density of states in S$_{k}$ at temperature $T_{k}$,\cite{GiazottoBergeret2} $\Gamma_{k}\sim 10^{-4}\Delta_{k}(0)$ and $k=1,2$. This term is peculiar to the Josephson effect (so it vanishes if one or both of the superconducting energy gaps are null) and originates from tunneling processes through the junctions involving both Cooper pairs and quasiparticles.\cite{MakiGriffin,Guttman} Depending on $\varphi$, $J_{\rm int}$ can flow in opposite direction with respect to that imposed by the thermal gradient, but the total heat current still flows from the hot to the cold reservoir, preserving the second principle of thermodynamics. This was experimentally demonstrated in Ref. \citenum{GiazottoNature}.

In the following, we shall assume, for clarity, that $\delta=\Delta_2/\Delta_1\leq 1$. Figure \ref{Fig2}(a) shows the behavior of $J_{fw}$ and $J_{rev}$ vs. $T_{\rm hot}$ for $T_{\rm cold}=0.01 T_{\rm c1}$ ($T_{\rm c1}$ being the critical temperature of S$_1$) and $\delta=0.75$. The difference between the heat currents is particularly evident for $\varphi=\pi$: in the forward configuration, $J_{\rm fw}$ exhibits a sharp peak at $T_{\rm hot}\simeq 0.77 T_{\rm c1}$, due to the matching of singularities in superconducting DOSs when $\Delta_1(T_{\rm hot})=\Delta_2(T_{\rm cold})$. At higher values of $T_{\rm hot}$, $\Delta_1(T_{\rm hot})<\Delta_2(T_{\rm cold})$ and the energy transmission through the junction is reduced, leading to a negative thermal differential conductance region, which is further enhanced by the gradual suppression of $J_{\rm int}$ as $T_{\rm hot}$ approaches $T_{\rm c1}$. In the reverse configuration, instead, $J_{\rm rev}$ presents just one cusp caused by the vanishing of $J_{\rm int}$ when $T_{\rm hot}=T_{\rm c2}=0.75 T_{\rm c1}$. Then, the rectification efficiency is maximum when the aforementioned points are perfectly aligned, i.e., $R_{\rm opt}\simeq 7.5$ for $\delta=0.75$ and $T_{\rm hot}\simeq 0.77 T_{\rm c1}$ at $T_{\rm cold}=0.01 T_{\rm c1}$, as shown in Figs. \ref{Fig2}(b) and \ref{Fig2}(c). As $T_{\rm cold}$ is increased, $R_{\rm opt}$ decreases monotonically to $\sim 1.3$ at $T_{\rm cold}=0.7 T_{\rm c1}$, then it becomes $<1$ and finally reaches the identity for $T_{\rm cold}=T_{\rm c1}$ (data not shown). More interestingly, the rectification efficiency can be finely tuned by controlling the phase bias across the junction. As a matter of fact, if $\varphi =0$, $\mathcal{R}$ is not only reduced by one order of magnitude but it also inverts the favored rectification regime, as displayed by Fig. \ref{Fig2}(d). This effect can be readily understood by observing the trend of $J_{fw}$ and $J_{rev}$ vs. $T_{\rm hot}$ for $\varphi=0$ [see Fig. \ref{Fig2}(a)]. In this case, the negative sign in front of $J_{\rm int}$ in Eq. \ref{int} 
perfectly cancels the singularity-matching peak in the forward configuration.

To conclude the analysis of the intrinsic properties of NIS and S$_1$IS$_2$ junctions, it is worthwhile to summarize the conditions necessary to achieve heat rectification: (1) two \textit{different} DOSs must be tunnel coupled (i.e., $\Delta_1\neq \Delta_2$) and (2) at least one of them must be strongly temperature dependent in the range of operation. Furthermore, a fairly large temperature gradient is needed, since heat rectification is absent in the linear response regime, i.e., when $\delta T=T_{\rm hot}-T_{\rm cold}\ll \overline{T}=(T_{\rm hot}+T_{\rm cold})/2$.\cite{MartinezAPL}
\section{NININ chain}

The rectifying performance of the discussed tunnel junctions can be largely improved if they are combined in an asymmetric chain of tunnel-coupled electrodes that can exchange energy with an independent phonon bath residing at temperature $T_{\rm bath}$. We stress that we are concerned with the heat current carried by electrons only. We assume that lattice phonons present in the entire structure are thermalized, with substrate phonons residing at $T_{\rm bath}$, and are therefore not responsible for any heat transport. This assumption is expected to hold because the Kapitza resistance between thin metallic films and the substrate is vanishingly small at low temperatures.\cite{GiazottoNature,MartinezNature,MartinezArxiv,Wellstood}

We shall begin with a simple N$_1$IN$_2$IN$_3$ chain,\cite{FornieriAPL} which, as we will show, represents a fundamental element to understand the working principles of chain diodes. N$_1$ and N$_3$ act as thermal reservoirs and are used to impose a temperature gradient across the device. Since $\mathcal{R}$ is defined under the condition of equal temperature bias, these electrodes must be identical and equally coupled to the phonon bath. Thus, $T_{\rm bath}$ now plays the role embodied by $T_{\rm cold}$ in the single junction analysis. N$_2$ is connected to N$_1$ and N$_3$ by means of two tunnel junctions characterized by resistances $R_1$ and $R_2$, respectively. For simplicity, we assume $R_1$ equal to 1 k$\Omega$ and consider only the parameter $r=R_2/R_1\geq 1$, accounting for the asymmetry of the chain. N$_2$ is the core of the diode and controls the heat flow from one reservoir to the other by releasing energy to the phonon bath. In the forward configuration, the electronic temperature of N$_1$ is set to $T_{\rm hot}>T_{\rm bath}$, leading to electronic thermal currents (see Eq. \ref{JNIN}) $J_{\rm 2,fw}$ and $J_{\rm fw}$ flowing into N$_2$ and N$_3$, respectively. On the contrary, in the reverse configuration the electronic temperature of N$_3$ is set to $T_{\rm hot}$, generating heat currents $J_{\rm 2,rev}$ and $J_{\rm rev}$ flowing into N$_2$ and N$_1$, respectively.

Besides electronic thermal currents, we must take into account the heat exchanged by electrons in the metal with lattice phonons:\cite{MartinezNature,Maasilta}
\begin{equation}
J_{\mathrm{N,ph}}(T,T_\mathrm{bath})=\Sigma \mathcal{V} (T^n-T_\mathrm{bath}^n).\label{Jqpph}
\end{equation}
Here $\Sigma$ is the material-dependent quasiparticle-phonon coupling constant, $\mathcal{V}$ is the volume of the electrode and $n$ is the characteristic exponent of the material. In this work we will consider two materials that are commonly exploited to realize N electrodes in nanostructures, i.e., copper (Cu) and manganese-doped aluminum (AlMn). The former is typically characterized by $\Sigma_{\rm Cu}=3\times 10^9$ WK$^{-5}$m$^{-3}$ and $n_{\rm Cu}=5$,\cite{GiazottoRev,GiazottoNature} while the latter exhibits $\Sigma_{\rm AlMn}=4\times 10^9$ WK$^{-6}$m$^{-3}$ and $n_{\rm AlMn}=6$.\cite{MartinezNature,MartinezArxiv,Maasilta} Furthermore, we assume that all the electrodes of the chain have a volume $\mathcal{V}_{\rm N}= 1 \times 10^{-20}$ m$^{-3}$.
\begin{figure*}[t]
\centering
\includegraphics[width=0.7\textwidth]{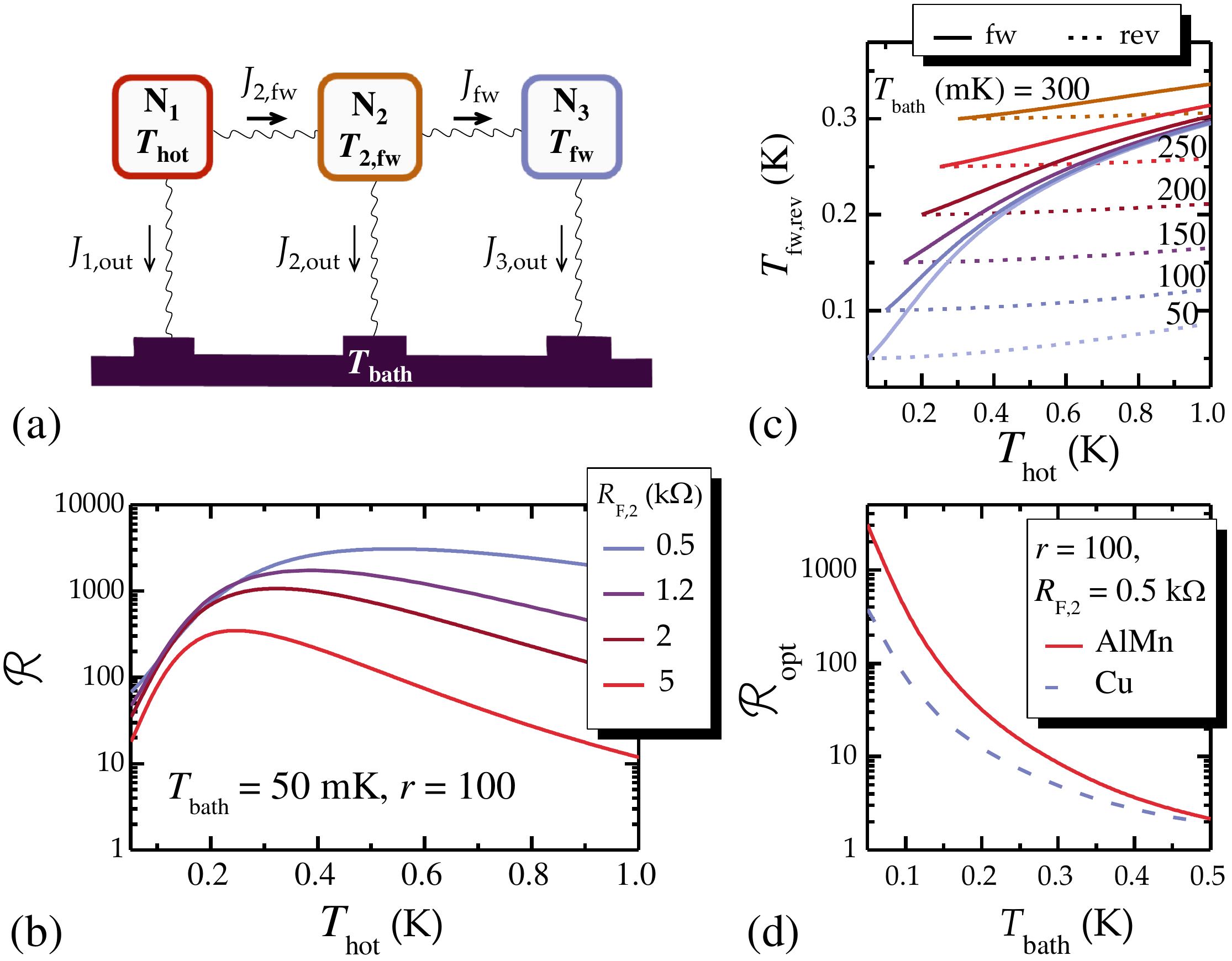}
\caption{(a) Thermal model for the NININ chain in the forward thermal bias configuration, which is characterized by $T_{\rm hot}> T_{\rm 2,fw}>T_{\rm fw}>T_{\rm bath}$. The most efficient design for the diode consists in tunnel-coupling a cold finger F$_2$ to N$_2$ and in minimizing the effect of the quasiparticle-phonon coupling in all the electrodes composing the chain (see text). (b) $\mathcal{R}$ vs. $T_{\rm hot}$ for different values of the cold finger resistance $R_{\rm F,2}$ at $T_{\rm bath}=50$ mK. (c) Diode's output temperatures $T_{\rm fw}$ (solid lines) and $T_{\rm rev}$ (dashed lines) vs. $T_{\rm hot}$ at different values of $T_{\rm bath}$. (d) $\mathcal{R}_{\rm opt}$ vs. $T_{\rm bath}$ for two different N materials. In panel (c) and (d) we set $R_{\rm F,2}=500\;\Omega$. All the results have been obtained for $r=100$.}
\label{Fig3}
\end{figure*}
Equations \ref{JNIN} and \ref{Jqpph} can be used to elaborate a thermal model accounting for heat transport through the device. 
The model is sketched in Fig. \ref{Fig3}(a) and describes the forward temperature bias configuration, in which the electrodes of the chain reside at temperatures $T_{\rm hot}> T_{\rm 2,fw}> T_{\rm fw}>T_{\rm bath}$. Here, $T_{\rm 2,fw}$ and $T_{\rm fw}$ represent the electronic temperatures of N$_2$ and N$_3$, respectively.
The terms $J_{\rm 2,fw}=J_{\rm NIN}(T_{\rm hot}, T_{\rm 2,fw})$ and $J_{\rm fw}=J_{\rm NIN}(T_{\rm 2,fw}, T_{\rm fw})$ account for the heat transferred from N$_1$ to N$_3$. The reservoirs can release energy to the phonon bath by means of $J_{\rm 1,out}$ and $J_{\rm 3,out}$. Photon-mediated thermal transport, \cite{MeschkeNature,Schmidt,Pascal} owing to poor impedence matching, as well as pure phononic heat currents are neglected in our analysis.\cite{GiazottoNature,MartinezNature,MartinezArxiv} We can now write a system of energy-balance equations that account for the detailed thermal budget in N$_2$ and N$_3$ by setting to zero the sum of all the incoming and outgoing heat currents:
\begin{align}
&J_\mathrm{2,fw}(T_\mathrm{hot},T_\mathrm{2,fw})-J_{\rm fw}(T_\mathrm{2,fw},T_{\rm fw})-J_\mathrm{2,out}(T_\mathrm{2,fw},T_\mathrm{bath})=0 \notag \\ 
&J_{\rm fw}(T_\mathrm{2,fw},T_{\rm fw})-J_\mathrm{3,out}(T_{\rm fw},T_\mathrm{bath})=0\label{eqs}.
\end{align}

Here $J_{\rm 2,out}(T_\mathrm{2,fw},T_\mathrm{bath})$ is the heat current that flows from N$_2$ to the phonon bath. In Eqs. \ref{eqs} we can set $T_{\rm hot}$ and $T_{\rm bath}$ as independent variables and calculate the resulting $T_{\rm 2,fw}$ and $T_{\rm fw}$. Another system of energy-balance equations can be written and solved for the reverse configuration,\cite{reverse} in which N$_2$ and N$_1$ reach electronic temperatures $T_{\rm 2,rev}$ and $T_{\rm rev}$, respectively. Finally, we can extract the values of $J_{\rm fw}$ and $J_{\rm rev}$, thereby obtaining $\mathcal{R}$. 

In order to grasp the essential physics underlying this device, it is instructive to consider an ideal case, in which N$_1$, N$_2$ and N$_3$ can release energy to the phonon bath only through three tunnel-coupled N electrodes, acting as thermalizing cold fingers\cite{FornieriAPL} and labeled as F$_1$, F$_2$ and F$_3$, respectively. We assume that electrons in F$_1$, F$_2$ and F$_3$ are strongly coupled to the phonon bath, so they reside at $T_{\rm bath}$. As we shall argue, F$_2$ is fundamental to achieve a high rectification performance and is characterized by a resistance $R_{\rm F,2}$, while F$_1$ and F$_3$ must have the same tunnel junction resistance $R_{\rm F,1}$. We ignore at this stage contributions due to the  direct quasiparticle-phonon coupling (see Eq. \ref{Jqpph}) in N$_1$, N$_2$ and N$_3$, in order to obtain simple analytic results for the rectification efficiency. As a matter of fact, this ideal device can be easily described by substituting $J_\mathrm{2,out}(T_\mathrm{2,fw},T_\mathrm{bath})=J_{\rm NIN}(T_{\rm 2,fw}, T_{\rm bath})$ and $J_\mathrm{3,out}(T_{\rm fw},T_\mathrm{bath})=J_{\rm NIN}(T_{\rm fw}, T_{\rm bath})$ in Eqs. \ref{eqs}. Remarkably, the resulting expression for the rectification efficiency does not depend on $T_{\rm hot}$ nor $T_{\rm bath}$, but only on the resistances of the tunnel junctions defining our NININ chain:
\begin{equation}
\mathcal{R}=\frac{R_1(R_2+R_{\rm F,2})+R_2(R_{\rm F,1}+R_{\rm F,2})+R_{\rm F,1}R_{\rm F,2}}{R_{\rm F,2}(R_2+R_{\rm F,1})+R_1(R_2+R_{\rm F,1}+R_{\rm F,2})}.\label{Rideal}
\end{equation}
Equation \ref{Rideal} is particularly transparent in two limit cases: first, if $R_{\rm F,2}\rightarrow\infty$, $\mathcal{R}\rightarrow 1$ for every value of $r$. This clearly demonstrates that no rectification is achievable if the central island of the chain is not coupled to the phonon bath, regardless the asymmetry of the device. This result holds true for a general design of a NININ chain, provided that N$_1$ and N$_3$ are identically coupled to the phonon bath.\cite{FornieriAPL}

On the other hand, if $R_{\rm F,1}\rightarrow\infty$, Eq. \ref{Rideal} is simplified as follows:
\begin{equation}
\mathcal{R}_{\rm no-ph}=\frac{R_2+R_{\rm F,2}}{R_1+R_{\rm F,2}}\label{Rasympt}.
\end{equation}
This expression summarizes the conditions needed to obtain an highly-efficient NININ diode: (1) the thermal symmetry of the device must be broken ($r\gg1$) and (2) N$_2$ must be very well coupled to the phonon bath with respect to N$_1$ and N$_3$. A natural trade off arising from the fulfilment of these conditions is the reduction of the thermal efficiency of the device $\eta=J_{\rm fw}/J_{\rm 2,fw}\leq 1$, i.e., the fraction of energy that is actually transferred from one reservoir to the other in the transmissive regime.\cite{FornieriAPL} Nevertheless, the device parameters can be designed in order to maximize the required performance in terms of the global efficiency $\mathcal{R}\eta$,\cite{FornieriAPL} which can range from 0 to $\infty$, since $\mathcal{R}$ has no upper limit. Finally, this ideal case demonstrates that such a thermal rectifier can operate efficiently also in the regime for $T_{\rm hot}\rightarrow T_{\rm bath}$. 

In a more realistic case, every electrode in the chain can exchange energy with the phonon bath by means of the quasiparticle-phonon coupling, which introduces a temperature dependence in the rectification efficiency. 
It is then useful to rewrite $\mathcal{R}\gg 1$ as a general condition for temperatures:
\begin{equation}
\mathcal{R}=\frac{\overline{T}_{\rm fw} \delta T_{\rm fw}}{r \overline{T}_{\rm rev}\delta T_{\rm rev}}\gg 1,\label{rect}
\end{equation}
where $\delta T_{\rm fw (rev)}= T_{\rm 2, fw (rev)}-T_{\rm fw (rev)}$ and the mean temperatures $\overline{T}_{\rm fw (rev)}=(T_{\rm 2,fw (rev)}+T_{\rm fw (rev)})/2$. In order to maximize $\mathcal{R}$, in the reverse configuration the electronic temperatures of N$_1$ and N$_2$ must be similar and close to the lowest temperature in the system, i.e., $T_{\rm bath}$. This can be easily done by setting $r>1$ and by coupling N$_2$ to the phonon bath. On the other hand, the coupling between N$_2$ and the phonon bath should have a limited impact on heat transport in the forward configuration. From Eq. \ref{Jqpph}, it appears evident that $J_{\rm N,ph}$ is not well-suited to satisfy the aforementioned requirements, since it is not much effective at low temperatures, while it becomes strongly invasive at high temperatures. Thus, it turns out that the most efficient design for a NININ chain diode is obtained by coupling just one thermalizing cold finger F$_2$ to N$_2$ and by minimizing the impact of the quasiparticle-phonon coupling in all the electrodes of the chain.\cite{FornieriAPL}

Then, we can set $J_\mathrm{2,out}(T_{\rm 2,fw},T_\mathrm{bath})=J_{\rm N,ph}(T_{\rm 2,fw},T_\mathrm{bath})+J_{\rm NIN}(T_{\rm 2,fw},T_\mathrm{bath})$ and  $J_\mathrm{3,out}(T_{\rm fw},T_\mathrm{bath})=J_{\rm N,ph}(T_{\rm fw},T_\mathrm{bath})$ in the energy-balance equations. Figure \ref{Fig3}(b) shows $\mathcal{R}$ as a function of $T_{\rm hot}$ for different values of $R_{\rm F,2}$ at $T_{\rm bath}=50$ mK. Remarkably, an optimal rectification efficiency\cite{ropt} $\mathcal{R}_{\rm opt}\sim 3000$ is obtained for $R_{\rm F,2}=500$ $\Omega$ and $r=100$. This configuration generates a maximum global efficiency $[\mathcal{R}\eta]_{\rm max}\simeq 8$, which can be optimized up to a value of about 9.6 by setting $R_{\rm F,2}=1.2$ k$\Omega$. We can also notice that for $T_{\rm hot}\rightarrow T_{\rm bath}$, $\mathcal{R}$ approaches $\mathcal{R}_{\rm no-ph}$, which represents an asymptotic value for the rectification efficiency at low temperatures, i.e., when the quasiparticle-phonon coupling is almost uneffective. 

Figure \ref{Fig3}(c) displays the rectifier's output temperatures $T_{\rm fw}$ and $T_{\rm rev}$ vs. $T_{\rm hot}$ for $r=100$ and $R_{\rm F}=500$ $\Omega$ at different values of $T_{\rm bath}$. The results point out a maximum difference of $\sim 200$ mK between the forward and reverse configurations at $T_{\rm bath}=50$ mK. Additionally, the behavior of output temperatures clearly highlights the effectiveness of F$_2$ in keeping $T_{\rm rev}$ close to $T_{\rm bath}$, while  $J_{\rm N,ph}$ strongly limits the increasing rate of $T_{\rm fw}$ as $T_{\rm hot}$ and $T_{\rm bath}$ become larger. 

Finally, Fig. \ref{Fig3}(d) confirms the negative effect of the quasiparticle-phonon coupling on the performance of the diode. As a matter of fact, the behavior of $R_{\rm opt}$ as a function of $T_{\rm bath}$ shows that the $T^5$ dependence of $J_{\rm N,ph}$ in Cu is extremely detrimental in the forward configuration and can reduce the rectification efficiency up to a factor of 10.

\begin{figure*}[t]
\centering
\includegraphics[width=0.8\textwidth]{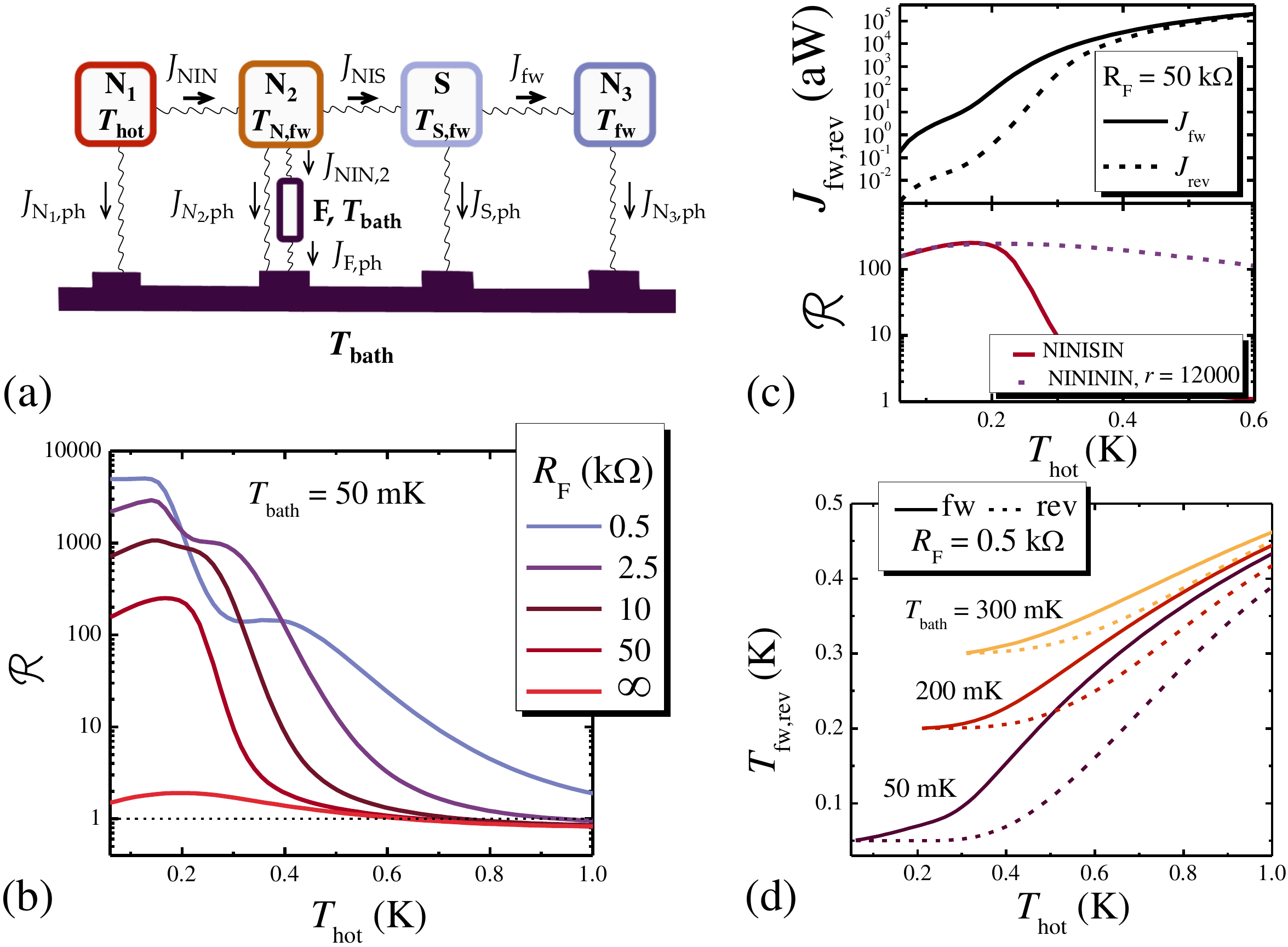}
\caption{(a) Thermal model for the NINISIN chain in the forward thermal bias configuration, characterized by $T_{\rm hot}> T_{\rm N,fw}>T_{\rm S,fw}>T_{\rm fw}>T_{\rm bath}$. (b) $\mathcal{R}$ vs. $T_{\rm hot}$ for different values of the cold finger resistance $R_{\rm F}$ at $T_{\rm bath}=50$ mK.  (c) The upper panel shows $J_{\rm fw}$ (solid line) and $J_{\rm rev}$ (dashed line) vs. $T_{\rm hot}$ for the NINISIN chain. The corresponding $\mathcal{R}$ is represented by the solid line in the lower panel, while the dashed line stands for the rectification efficiency of a NINININ chain with a resistance asymmetry $r=12000$. In both the panels we set $R_{\rm F}=50$ k$\Omega$ and $T_{\rm bath}=50$ mK.  (d) Diode's output temperatures $T_{\rm fw}$ (solid lines) and $T_{\rm rev}$ (dashed lines) vs. $T_{\rm hot}$ at different values of $T_{\rm bath}$ for $R_{\rm F}=0.5$ k$\Omega$.}
\label{Fig4}
\end{figure*}

\section{NINISIN chain}

Despite its conceptual simplicity and potential insensitivity to magnetic fields, the NININ chain diode might be demanding from a fabrication point of view because of the required high asymmetry between $R_1$ and $R_2$. A good alternative is represented by a N$_1$IN$_2$ISIN$_3$ chain, in which all the tunnel-junction normal-state resistances $R_1$, $R_2$ and $R_3$ are equal and the thermal symmetry is broken by the presence of the energy gap in the superconducting density of states. At low temperatures, the latter plays a role equivalent to the one embodied by the largest tunnel-junction resistance in a NININ chain. In the following analysis, we will consider an S electrode made of aluminum with a critical temperature $T_{\rm c}=1.4$ K. 

Emulating the most efficient design for a NININ chain, also in this case the N$_2$ electrode can be coupled to the phonon bath by means of a cold finger F residing at $T_{\rm bath}$. On the other hand, it is worthwhile to note that a simple NISIN chain would not represent an efficient design because $\Delta(T)$ would prevent the central electrode to release energy to F in an efficient way, especially at low temperatures [see Fig. \ref{Fig1}(a)]. Figure \ref{Fig4}(a) sketches the thermal model accounting for thermal transport through the device in the forward configuration, characterized by $T_{\rm hot}>T_{\rm N,fw}>T_{\rm S,fw}>T_{\rm fw}>T_{\rm bath}$. Here, $T_{\rm N,fw}$ and $T_{\rm S,fw}$ label the electronic temperatures of N$_2$ and S, respectively. We assume that all the electrodes have the same volume $\mathcal{V}=1\times 10^{-20}$ m$^{-3}$ and $R_1=R_2=R_3=1$ k$\Omega$. Heat is transferred from N$_1$ to N$_3$ by means of the heat currents $J_{\rm NIN}(T_{\rm hot},T_{\rm N,fw})$, $J_{\rm NIS}(T_{\rm N,fw},T_{\rm S,fw})$ and $J_{\rm fw}=J_{\rm NIS}(T_{\rm S,fw},T_{\rm fw})$. Electrode N$_2$ can release energy to the phonon bath via $J_{\rm 2,out}=J_{\rm NIN}(T_{\rm N,fw},T_{\rm bath})+J_{\rm N,ph}(T_{\rm N,fw},T_{\rm bath})$, while N$_3$ is coupled to $T_{\rm bath}$ through $J_{\rm N,ph}(T_{\rm fw},T_{\rm bath})$. Finally, S is affected by the quasiparticle-phonon coupling as well, which can be written as:\cite{Timofeev}
\begin{align}
&J_{\mathrm{S,ph}}(T,T_\mathrm{bath})=-\frac{\Sigma \mathcal{V}}{96 \zeta (5)k_\mathrm{B}^5}\int_{-\infty}^{\infty} dE E \int_{-\infty}^{\infty} d\mathrm{\epsilon}\mathrm{\epsilon}^2 \mathrm{sgn}(\mathrm{\epsilon}) \notag \\
&\times L_{E,E+\mathrm{\epsilon}}\left[\mathrm{coth}\left(\frac{\mathrm{\epsilon}}{2k_\mathrm{B}T_\mathrm{bath}}\right)(f_E^{\mathrm{(1)}}-f_{E+\mathrm{\epsilon}}^{\mathrm{(1)}})-f_E^{\mathrm{(1)}}f_{E+\mathrm{\epsilon}}^{\mathrm{(1)}}+1\right], \label{superph}
\end{align}
where $f_E^\mathrm{(1)}=f(-E,T)-f(E,T)$ and the phonons are assumed to be in thermal equilibrium with occupation $n(\mathrm{\epsilon},T_\mathrm{bath})=[\mathrm{exp}(\frac{\mathrm{\epsilon}}{k_\mathrm{B}T_\mathrm{bath}})-1]^{-1}$. The factor $L_{E,E'}=\mathcal{N}(E,T)\mathcal{N}(E',T)\left[1-\frac{\mathrm{\Delta^2(T)}}{EE'} \right]$. In our case, we assume $\Sigma_\mathrm{Al}=$ 0.3 $\times$ 10$^9$ WK$^{-5}$m$^{-3}$.\cite{GiazottoRev} With all these ingredients we can write and solve the energy balance equations for the forward and reverse configuration, thus allowing us to analyze the operation of the diode vs. $T_{\rm hot}$ and $T_{\rm bath}$ for different parameters. 

Figure \ref{Fig4}(b) displays the behavior of $\mathcal{R}$ as a function $T_{\rm hot}$ for different values of the resistance $R_{\rm F}$ between F and N$_2$ at $T_{\rm bath}=50$ mK. At low values of $T_{\rm hot}$ the rectification efficiency can reach values as high as 5000 for $R_{\rm F}=0.5$ k$\Omega$, but for increasing temperatures it dramatically drops and approaches unity for $T_{\rm hot}>1$ K. This trend can be easily understood by first focusing on the case of $R_{\rm F}>10$ k$\Omega$. The upper panel of Fig. \ref{Fig4}(c) shows the diode's output currents $J_{\rm fw (rev)}$ vs. $T_{\rm hot}$ for $R_{\rm F}=50$ k$\Omega$ and directly explains the $\mathcal{R}$ curve (solid line) reported in the lower panel. As already noticed at the end of Sect. \ref{NISsec}, for sufficiently high thermal bias an S electrode loses its effectiveness as a thermal bottleneck, as can be noticed from the change in the derivative of the output currents for $T_{\rm hot}\gtrsim 0.2$ K in both the forward and reverse configurations. This change of behavior produces a maximum in the rectification efficiency at $T_{\rm hot}\simeq 0.15$ K, after which $\mathcal{R}$ rapidly decreases. Then, the first part of the curve can be compared to the efficiency (dashed line) of an equivalent NINININ chain with identical volumes of electrodes and a resistance asymmetry $r=R_3/R_1=12000$, which is about the value of $\Delta(0)/\Gamma=10^4$. On the other hand, for $R_{\rm F}<10$ k$\Omega$, $T_{\rm S,rev}$ can be significantly higher than $T_{\rm S,fw}$ at the same value of $T_{\rm hot}$. In particular, for $R_{\rm F}=0.5$ k$\Omega$ in the reverse configuration S loses its properties of thermal bottleneck for $T_{\rm hot}\gtrsim 0.15$ K, while in the forward configuration $\Delta$ blocks efficiently the thermal flux up to $T_{\rm hot}\simeq 0.27$ K. This discrepancy in the "breaking points" of the S electrode is reflected in the behavior of $\mathcal{R}$, which exhibits a shoulder after the first maximum [see Fig.\ref{Fig4}(b)]. Furthermore, it is worth noting that for $R_{\rm F}>2.5$ k$\Omega$ the device can operate also in the reverse regime of rectification at sufficiently high $T_{\rm hot}$, reaching values of $\mathcal{R}\simeq 0.8$, as expected for a simple NIS junction (see Sect. \ref{NISsec}).

As $T_{\rm bath}$ is increased, the maximum value of $\mathcal{R}$ in the forward regime becomes strongly suppressed due to the invasive enhancement of the quasiparticle-phonon coupling. This can be indirectly observed in Fig.\ref{Fig4}(d), which shows the diode's output temperatures $T_{\rm fw (rev)}$ vs. $T_{\rm hot}$ for $R_{\rm F}=0.5$ k$\Omega$ at three selected values of $T_{\rm bath}$. At $T_{\rm bath}=50$ mK we obtain a maximum temperature difference $\delta T=T_{\rm fw}-T_{\rm rev}$ of about 110 mK, while at higher $T_{\rm bath}$ it rapidly decreases in conjunction with the derivative of $T_{\rm fw (rev)}$.

Remarkably, the global efficiency of this device at $T_{\rm bath}=50$ mK is comparable to the NININ case and can be optimized to $[\mathcal{R}\eta]_{\rm max}\simeq 7.5$ by setting $R_{\rm F}=7.5$ k$\Omega$.

Finally, it is also possible to use a different superconductor with a larger $\Delta$ (such as vanadium or niobium) to implement the S electrode of the chain. This can significantly increase the range of $T_{\rm hot}$ where the superconductor can act as a large thermal resistance, thus providing $\mathcal{R}\gg 1$. Nevertheless, the performance of the device vs. $T_{\rm bath}$ will not be significantly improved, since it is mainly determined by the disruptive effect of the quasiparticle-phonon coupling in N electrodes.  

The effectiveness of this kind of device has been recently proven in the experiment reported in Ref. \citenum{MartinezArxiv}.
\section{NINIS$_2$IS$_1$IN chain}
\begin{figure*}[t]
\centering
\includegraphics[width=0.8\textwidth]{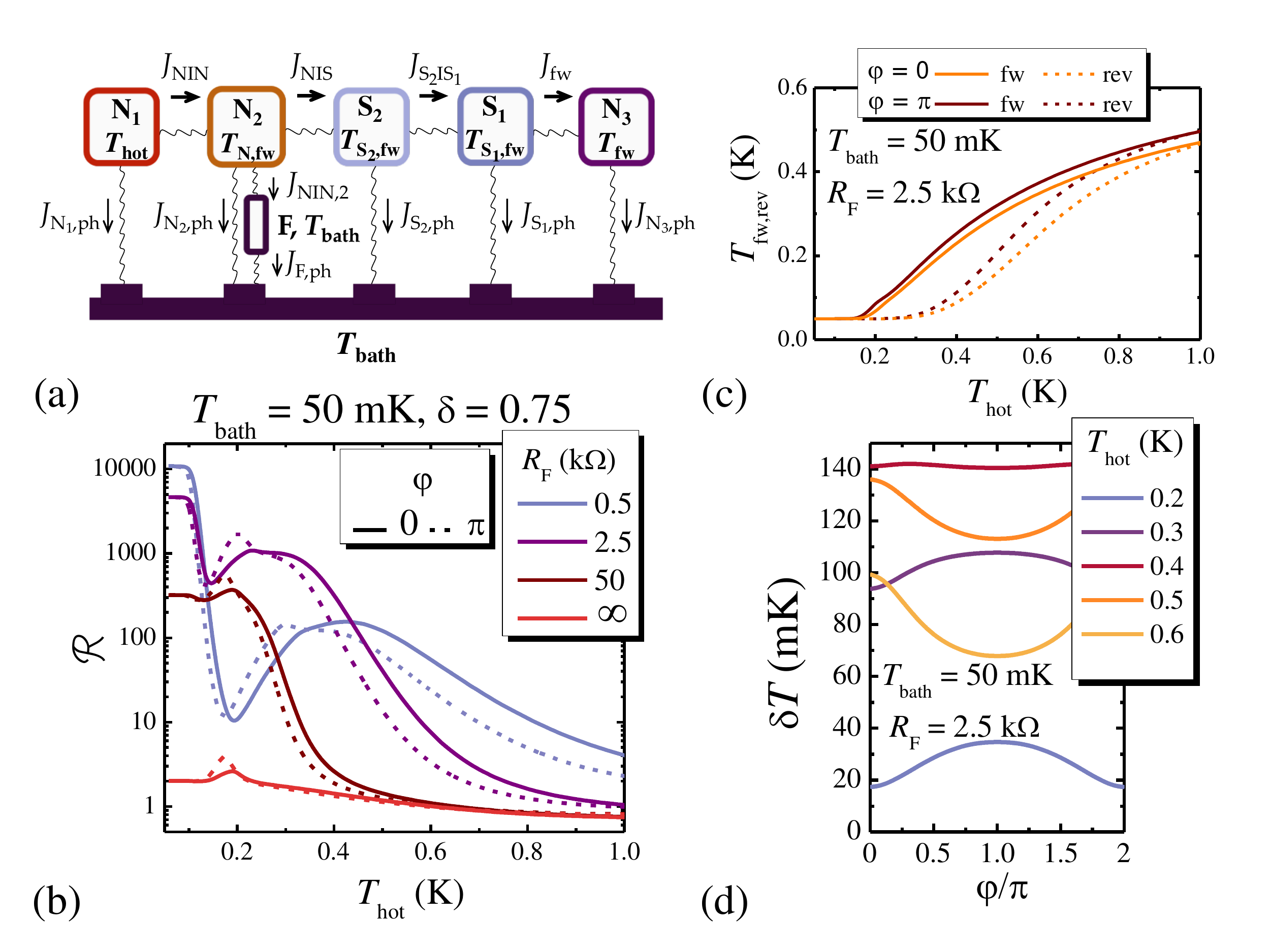}
\caption{(a) Thermal model for the NINIS$_2$IS$_1$IN chain in the forward thermal bias configuration, characterized by $T_{\rm hot}> T_{\rm N,fw}>T_{\rm S_2,fw}>T_{\rm S_1,fw}>T_{\rm fw}>T_{\rm bath}$. (b) $\mathcal{R}$ vs. $T_{\rm hot}$ for different values of the cold finger resistance $R_{\rm F}$ at $T_{\rm bath}=50$ mK. Solid lines correspond to the case in which $\varphi=0$, while dashed lines represent $\mathcal{R}$ when $\varphi=\pi$. (c) Diode's output temperatures $T_{\rm fw}$ (solid lines) and $T_{\rm rev}$ (dashed lines) vs. $T_{\rm hot}$ for $R_{\rm F}=2.5$ k$\Omega$ and for $\varphi=0,\pi$ at $T_{\rm bath}=50$ mK. (d) $\delta T = T_{\rm fw}-T_{\rm rev}$ vs. $\varphi$ for different values of $T_{\rm hot}$ at $T_{\rm bath}=50$ mK and for $R_{\rm F}=2.5$ k$\Omega$. }
\label{Fig5}
\end{figure*}

In previous works,\cite{MartinezAPL,MartinezRev} we analyzed theoretically the performance of a S$_1$IS$_2$ junction tunnel-coupled to two N electrodes acting as thermal reservoirs. This thermal diode exhibits a remarkable phase-modulation of the rectification efficiency for $T_{\rm c1}=1.4$ K and $\delta=0.75$ at $T_{\rm bath}=10$ mK: the best performance is reached at $T_{\rm hot}=1.15$, where $\mathcal{R}$ can be varied from a maximum of $\sim 4.4$ (if $\varphi= \pi$) to a minimum of $\sim 0.7$ (if $\varphi=0$).\cite{MartinezAPL} Thus, the behavior of a NIS$_1$IS$_2$IN chain can be fully tuned by varying the phase difference between S$_1$ and S$_2$, allowing to invert the direction of the rectification regime. Nevertheless, this device has some experimental drawbacks, since its optimal working point requires a large bias temperature, close to the critical temperature of aluminum. As a consequence, superconducting thermometers used to detect $T_{\rm hot}$, $T_{\rm fw}$ and $T_{\rm rev}$ (as in Refs. \citenum{GiazottoNature,MartinezNature,MartinezArxiv}) should be realized with a superconductor S$_3$, characterized by an energy gap $\Delta_3 \gg \Delta_1,\Delta_2$, thus complicating the experimental design of the rectifier. Therefore, we can think of employing an additional N electrode connected to the phonon bath by means of a cold finger in order to improve $\mathcal{R}_{\rm opt}$ and to move the optimal working point of the device towards lower values of $T_{\rm hot}$. As we will argue, the proposed design offers the possibility to increase $\mathcal{R}_{\rm opt}$ by more than 3 orders of magnitude with respect to the NIS$_1$IS$_2$IN chain diode and to demonstrate experimentally the phase modulation of $\mathcal{R}$ in a relatively simpler way. However, it prevents us inverting the rectification regime through the phase modulation. 

We consider a N$_1$IN$_2$IS$_2$IS$_1$IN$_3$ chain, in which N$_2$ is tunnel-coupled to a cold finger F residing at $T_{\rm bath}$, S$_1$ has a critical temperature $T_{\rm c1}=1.4$ K and $\delta=\Delta_2/\Delta_1=0.75$. This design results to be the one which maximizes the effect of a $\varphi$ variation on the rectification efficiency. Figure \ref{Fig5}(a) displays the thermal model accounting for thermal transport through the device in the forward configuration, characterized by $T_{\rm hot}>T_{\rm N,fw}>T_{\rm S_2,fw}>T_{\rm S_1,fw}>T_{\rm fw}>T_{\rm bath}$. Here, $T_{\rm N,fw}$, $T_{\rm S_1,fw}$ and $T_{\rm S_2,fw}$ label the electronic temperatures of N$_2$, S$_1$ and S$_2$, respectively. We assume that all the electrodes have the same volume $\mathcal{V}=1\times 10^{-20}$ m$^{-3}$ and all the tunnel junction have a normal-state resistance $R_1=R_2=R_3=R_4=1$ k$\Omega$. Heat is transferred from N$_1$ to N$_3$ by means of the heat currents $J_{\rm NIN}(T_{\rm hot},T_{\rm N,fw})$, $J_{\rm NIS}(T_{\rm N,fw},T_{\rm S_2,fw})$, $J_{\rm S_1IS_2}(T_{\rm S_2,fw},T_{\rm S_1,fw})$ and $J_{\rm fw}=J_{\rm NIS}(T_{\rm S_1,fw},T_{\rm fw})$. Electrode N$_2$ can release energy to the phonon bath via $J_{\rm 2,out}=J_{\rm NIN}(T_{\rm N,fw},T_{\rm bath})+J_{\rm N,ph}(T_{\rm N,fw},T_{\rm bath})$, while N$_3$ is coupled to $T_{\rm bath}$ through $J_{\rm N,ph}(T_{\rm fw},T_{\rm bath})$. Finally, S$_1$ and S$_2$ are affected by the quasiparticle-phonon coupling $J_{\rm S_{1,2},ph}(T_{\rm S_{1,2}fw},T_{\rm bath})$. In general, the junction S$_2$IS$_1$ can be phase biased through supercurrent injection or by applying an external magnetic flux.\cite{MartinezAPL} In order to achieve a complete modulation of $\mathcal{R}$, $\varphi$ must vary between 0 and $\pi$. This can be obtained by using a radio frequency superconducting quantum interference device (rf SQUID), but this would again require the use of a clean contact between the S$_2$IS$_1$ junction and a third superconductor S$_3$ (with $\Delta_3 \gg \Delta_1,\Delta_2$) in order to suppress heat losses.\cite{MartinezAPL,MartinezRev} A good alternative is represented by an asymmetric direct current SQUID formed by different Josephson junctions (JJs), of which one corresponds to the S$_2$IS$_1$ junction. The latter can be biased up to $\varphi=\pi$ if its Josephson inductance\cite{Tinkham} is the largest in the SQUID, i.e., its Josephson critical current is much lower than those of the other JJs. 

As done in previous cases, we can now write and solve the energy-balance equations accounting for the thermal budget in every electrode of the chain. Thus, we can obtain the behavior of the thermal diode vs. $T_{\rm hot}$, $T_{\rm bath}$ and $\varphi$ for different values of the parameters.

Figure \ref{Fig5}(b) shows $\mathcal{R}$ vs. $T_{\rm hot}$ for different values of the cold finger resistance $R_{\rm F}$ at $T_{\rm bath}=50$ mK. Solid lines represent the rectification efficiency for $\varphi=0$, while dashed lines stand for that obtained in the case of $\varphi=\pi$. The values of $\mathcal{R}$ provided here are comparable to those obtained with the NINISIN chain, but the phase dependence of $J_{\rm S_1IS_2}$ represents an additional ingredient that strongly influences the performance of the diode over a large part of the $T_{\rm hot}$ range. In particular, for $R_{\rm F}=0.5$ k$\Omega$ we can reach a maximum relative variation $[\mathcal{R}(0)-\mathcal{R}(\pi)]/\mathcal{R}(0)\simeq 77$\% at $T_{\rm hot}=0.13$ K. From the point of view of output temperatures, though, the phase tuning of the device is clearly detectable only for $T_{\rm hot}>0.2$ K. The effect of the phase biasing is especially visible on $T_{\rm rev}$, which is affected by a maximum variation of $\sim$ 60 mK at $T_{\rm hot}\simeq$ 0.6 K. At this temperature also the phase modulation of $\delta T=T_{\rm fw}-T_{\rm rev}$ exhibits the largest amplitude (of $\sim 30$ mK), as shown in Fig. \ref{Fig5}(d).

We finally notice that the global efficiency of a N$_1$IN$_2$IS$_2$IS$_1$IN$_3$ chain is comparable to those of other chain diodes, with a maximum $[\mathcal{R}\eta]_{\rm max} \sim 6$ for $R_{\rm F}=8$ k$\Omega$.
\section{Summary and final remarks}
In summary, we first reviewed the intrinsic properties of superconducting hybrid junctions as thermal rectifiers. NIS and S$_1$IS$_2$ junctions can reach an optimal rectification efficiency of $\sim 0.8$\cite{MartinezAPL,GiazottoBergeret} and $\sim 7.5$,\cite{MartinezAPL,MartinezRev} respectively. On the other hand, even though a simple NIN junction cannot provide heat rectification, it is possible to obtain $\mathcal{R}_{\rm opt}$ as high as 3000 by realizing an asymmetric NININ chain with the central island coupled to the phonon bath by means of a cold finger. The latter provides an efficient channel through which the diode can release energy in the non-transmissive temperature bias configuration.\cite{FornieriAPL} This design can be used also in a NINISIN chain, which offers the opportunity to boost the performance of a single NIS junction and to overcome the fabrication complexity imposed by the high resistance asymmetry required in a NININ chain. The superconducting energy gap plays the role of a thermal bottleneck at low temperatures, while for higher temperatures it can generate an inversion of the rectification regime, as experimentally proven in Ref. \citenum{MartinezArxiv}. Finally, we analyzed a NINIS$_2$IS$_1$IN chain as a testing ground for the experimental demonstration of a phase-tunable thermal diode, providing a larger rectification efficiency and requiring lower bias temperatures with respect to those obtained in previous theoretical proposals.\cite{MartinezAPL,MartinezRev}
These thermal rectifiers could be easily implemented by standard nanofabrication techniques \cite{MartinezArxiv} and, combined with heat current interferometers, \cite{GiazottoNature,MartinezNature} might become building blocks of coherent caloritronic nanocircuits.\cite{GiazottoNature,MartinezNature,MartinezRev} 
Moreover, they could be connected to electronic coolers and radiation detectors,\cite{GiazottoRev} offering the possibility to evacuate dissipated power in an unidirectional way. Finally, this technology might also have a potential impact in general-purpose cryogenic electronic microcircuitry, e.g., solid-state quantum information architectures.\cite{NielsenChuang}

\section{Acknowledgments}
The European Research Council under the European Union's Seventh Framework Programme (FP7/2007-2013)/ERC Grant Agreement No. 615187-COMANCHE  is acknowledged for partial financial support.




\begin{thebibliography}{99}
\bibitem{reverse} Energy-balance equations for the reverse configuration can be easily obtained by replacing $J_{\rm 2,fw}\Rightarrow J_{\rm 2,rev}$, $J_{\rm fw}\Rightarrow J_{\rm rev}$, $J_{\rm 3,out}\Rightarrow J_{\rm 1,out}$, $T_{\rm 2,fw}\Rightarrow T_{\rm 2,rev}$ and $T_{\rm fw}\Rightarrow T_{\rm rev}$.
\bibitem{ropt} In the case of chain diodes, $\mathcal{R}_{\rm opt}$ is defined as the rectification efficiency corresponding to the maximum value between $\mathcal{R}$ and $1/\mathcal{R}$ at a given $T_{\rm bath}$.
\end{thebibliography}
\end{document}